
\input phyzzx

\PHYSREV

\titlepage
\title{
Teleportation of quantum states
}
\author{Lev Vaidman}

\address{
School of
Physics and Astronomy \break
Raymond and Beverly Sackler Faculty of Exact Sciences \break
Tel-Aviv University, Tel-Aviv, 69978 ISRAEL}
\vskip 3 cm
hep-th/9305062
\vfill

\noindent
BITNET: LEV@TAUNIVM
\endpage

Bennett {\it et al.}\Ref\BEN{C.H. Bennett, G. Brassard, C. Crepeau,
R. Jozsa, A. Peres, and W.K. Wootters, Phys. Rev. Lett.,
{\bf70}, 1895 (1993).}
have shown how to transfer (``teleport") an unknown spin quantum
state by using
prearranged correlated quantum systems and transmission of
classical
information. I will show how their results can be obtained in the
framework of nonlocal measurements proposed by Aharonov and
Albert\rlap.\Ref\AA{ Y.~Aharonov and D.~Albert,
 Phys.  Rev. {\bf D21}, 3316 (1980); {\bf D24}, 359 (1981).}\
 I will generalize the latter
 to the teleportation of
 a quantum state of a system with continuous variables.

We call a measurement {\it
nonlocal} if it cannot
 be reduced to a set of local measurements.
An example is a measurement of a sum of spin components of
separated particles. The EPR-Bohm state of two spin-1/2 particles
employed by Bennett {\it et al.} can be verified using
two consecutive measurements: ${\sigma_1}_x +
{\sigma_2}_x$ and then ${\sigma_1}_y + {\sigma_2}_y$. If the outcomes
are $${\sigma_1}_x(t_1) + {\sigma_2}_x(t_1) = 0, ~~~~
{\sigma_1}_y(t_2) + {\sigma_2}_y(t_2) = 0,\eqno(1)$$
 where $t_2 > t_1$,  then, after time $t_2$, the system is in
the EPR-Bohm state.
Had we started at time $t < t_1$ with the EPR-Bohm state, we
would be certain to obtain the outcomes (1).

The method of Aharonov and Albert is applicable also to
measurements which are nonlocal not only in space but also in
time. It has been shown that any sums and modular sums of local
variables are measurable\rlap.\Ref\AAV{Y.~Aharonov, D.~Albert, and L.~Vaidman,
 Phys.  Rev.  {\bf D34}, (1986) 1805.}\
 In
particular, we can perform ``crossed" measurements of
 ${\sigma_1}_x(t_1) - {\sigma_2}_x(t_2)$ and
${\sigma_1}_y(t_2) - {\sigma_2}_y(t_1)$. If the outcomes are
$${\sigma_1}_x(t_1) - {\sigma_2}_x(t_2) = 0, ~~~~
{\sigma_1}_y(t_2) - {\sigma_2}_y(t_1) = 0,\eqno(2)$$
then we obtain {\it complete correlation} between the state
of particle 1 before $t_1$ and particle 2 after $t_2$. Thus,
we succeed in teleporting  the state of particle 1
to particle 2.
However, this is not good enough, since the nonlocal measurements might not
yield outcomes (2). In that case we destroy the state
without teleporting it. In order to obtain {\it reliable}
teleportation (such as the one suggested by Bennett {\it et al.})
we must measure, instead, the following nonlocal observables:
$$\bigl({\sigma_1}_x(t_1) - {\sigma_2}_x(t_2)\bigr){\rm mod4}, ~~~~
\bigl({\sigma_1}_y(t_2) - {\sigma_2}_y(t_1)\bigr){\rm
mod4}.\eqno(3)$$
The outcome 0 brings us to the
 previous
case. If, however, the outcome is 2 for a given axis, then we can
convert it to 0 by rotation of the coordinate frame of the second
particle (${\sigma_2}_x = - {\sigma_2}_{x'}$, for  {\it \^x}$' = -$
{\it\^x}). Thus,
 for any set of outcomes of the nonlocal measurements (3) we teleport the
spin state; in some cases it gets rotated but we know when and
how from the
results of the nonlocal measurements. We can complete, then, the
teleportation by appropriate rotation.

The Aharonov-Albert method for nonlocal  measurement
consists
of a preparation of an entangled state of the measuring device,
local interactions with separate parts, and local reading of
the
separate parts of the measuring device resulting in a set of numbers
obtained in the respective space-time locations of the parts of the
system. These numbers
represent classical information which must be transmitted for
completing
the teleportation. (In our example, the information tells us which
rotation must be performed). The initial entanglement of the
measuring device, which is the core of the method, may employ pairs of
 spin-1/2 particles\refmark{\AAV} in the EPR-Bohm state (see Sec.IV
of Ref. (\AAV)), making this
method a variation of the Bennett {\it et al.}
proposal\rlap;\Ref\FOT{The
former has a small advantage in that it yields two-way teleportation:
also the state of particle 2 is transmitted to particle 1}\
but it can also
employ a system with continues variables (see Sec. II of Ref. (\AAV).
Using this method we can perform nonlocal measurements of
continuous
variables, and consequently teleport the corresponding
quantum states.

Consider two similar systems located far away from each other and
described by continuous
variables $q_1$, $q_2$ and conjugate momenta $p_1$ and $p_2$. In order
to teleport a quantum state $\Psi(q_1)$ we will perform the following nonlocal
measurements, obtaining the outcomes $a$ and $b$,
$$q_1(t_1) - q_2(t_2) = a, ~~~~ p_1(t_2) - p_2(t_1) = b. \eqno(4)$$
These nonlocal ``crossed" measurements will result in correlation of
the state of particle 1 before $t_1$ and the state of particle 2
after $t_2$, thus teleporting the quantum
state to the second particle up to a shift of
$-a$ in $q$ and $-b$ in $p$.
 These shifts are known (after results of local
measurements are transmitted), and can easily be corrected even
when the state is unknown, thus completing a
 reliable teleportation of the
state $\Psi(q_1)$ to $\Psi(q_2)$. A generalization of the Bennett
{\it et al.} scheme to the case of a continuum is also possible; the
essential ingredients appear (in
another context) elsewhere\rlap.\Ref\AAA{Y.~Aharonov, D.~Albert, and
C.K. Au,
Phys. Rev.
Lett., {\bf 47}, (1981) 1029.}

It is a pleasure to thank Yakir Aharonov for very
helpful suggestions.  The research was supported in part by grant
425/91-1 of the the Basic Research Foundation (administered by the
Israel Academy of Sciences and Humanities).

\refout

 \end